\documentclass[aps,prl,twocolumn,groupedaddress,showpacs]{revtex4}

\usepackage{graphicx}
\usepackage{amsmath}

\begin{document}

\title{Thermodynamic properties of Bi$_2$Sr$_2$CaCu$_2$O$_8$ calculated from the electronic dispersion.}

\author{J.G. Storey$^1$, J.L. Tallon$^{1,2}$, G.V.M. Williams$^{2}$}

\affiliation{$^1$School of Chemical and Physical Sciences,
Victoria University, P.O. Box 600, Wellington, New Zealand}

\affiliation{$^2$MacDiarmid Institute, Industrial Research Ltd.,
P.O. Box 31310, Lower Hutt, New Zealand}

\date{\today}

\begin{abstract}
The electronic dispersion for Bi$_2$Sr$_2$CaCu$_2$O$_{8+\delta}$ has
been determined from angle-resolved photoelectron spectroscopy
(ARPES). From this dispersion we calculate the entropy and
superfluid density. Even with no adjustable parameters we obtain an
exceptional match with experimental data across the entire phase
diagram, thus indirectly confirming both the ARPES and thermodynamic
data. The van Hove singularity is crossed in the overdoped region
giving a distinctive linear-in-$T$ temperature dependence in the
superfluid density there.
\end{abstract}


%%\begin{multicols}{2}
\maketitle

The generic doping dependence of the thermodynamic, electrodynamic
and transport properties of high-$T_c$ superconductors remains a
puzzle despite many years of study. Their unusual behaviour is often
taken to be a signature of exotic physics yet it should be related
to the electronic energy-momentum dispersion, obtained for example
from angle-resolved photoemission spectroscopy (ARPES). These
studies indicate the presence of an extended van Hove saddle-point
singularity\cite{VHS3} situated at the (0,$\pi$) point together with
a normal-state pseudogap\cite{PG2} as common features in the
electronic band structure of the cuprate superconductors. The
pseudogap exhibits a reduction in the density of states (DOS) at the
Fermi level which is believed to develop into a fully nodal gap at
low temperature\cite{FERMIARCS2}.

In theories based on the so-called van Hove
scenario\cite{VHSCENARIO} the superconducting (SC) transition
temperature, $T_{c}$, is enhanced by the close proximity of a van
Hove singularity (vHs). These theories assume that the vHs sweeps
through the Fermi level at around optimal doping ($p=0.16$), indeed
causing the peak in $T_{c}(p)$. However, ARPES measurements of the
band structure of Bi-2201\cite{2201VHS} show that the vHs crosses
the Fermi level in the deeply overdoped side of the phase diagram.
For a bilayer cuprate like Bi-2212 the weak coupling between the
layers splits the bands near the $(\pi,0)$ points into an upper
antibonding band and a lower bonding band. ARPES measurements
performed on Bi-2212\cite{2212VHS} suggest that the antibonding band
vHs crosses at around $p=0.225$ where $T_{c}\approx 60$K i.e. near
the limit for overdoping in this material. The vHs crossing should
profoundly affect all physical properties.

In this work we have used the thermodynamic properties as a window
on the electronic structure to independently check the main
ARPES results. Using an ARPES-derived energy dispersion we have calculated the doping and
temperature dependence of the entropy and superfluid density of Bi-2212. All details for
our calculations are taken directly, and only, from ARPES in order
to determine the implications of this data. Our calculations confirm
the ARPES results, giving a consistent picture of the thermodynamic
and electrodynamic properties in terms of a proximate vHs.

We assume Fermi-liquid-like, mean-field, weak-coupling physics in
spite of indications, or expectations, to the contrary. In defense
of our approach (i) the thermodynamics at low $T$ is dominated by
the nodal regions of the Fermi surface where quasiparticles are well
defined and long lived; and (ii) the Wilson ratio relating spin
susceptibility to $S/T$ (where $S$ is the electronic entropy) is
almost exactly that for nearly free electrons across a wide range of
doping and temperature\cite{ENTROPYDATA2}. Moreover, the
non-mean-field BCS-like behaviour is generally inferred from the
unusual $2\Delta/k_BT_c$ ratio which grows with underdoping. But it
has always been our view\cite{OURWORK1}, and is now
confirmed\cite{TANAKA,LETACON} that the large and growing energy gap
used here is the ($\pi$,0) pseudogap, not the SC gap. Once the
pseudogap is properly included in the problem then $2\Delta/k_BT_c$
is well behaved. We do not treat fluctuations which are confined to
$T_c \pm 15$K\cite{SCFLUC} and are a minor embellishment.

For Bi-2212 we employ a 2D bilayer dispersion
$\epsilon_{\textbf{k}}$ provided by the authors of
ref.~\cite{2212VHS}, which was obtained from tight binding fits to
high-resolution ARPES data. The DOS per spin at energy $E$ is given
by
\begin{equation}
N(E)=N_k^{-1}\sum_{\textbf{k}}\delta\left(\epsilon_{\textbf{k}}-E\right)
\label{DOSEQ}
\end{equation}
The entropy per mole $S$ for weakly interacting fermions is given by\cite{PARKS}
\begin{equation}
S=-2R\int{[f\ln{f}+(1-f)\ln{(1-f)}]N(E)dE}
\label{ENTROPYEQ}
\end{equation}
where $f$ is the Fermi-Dirac distribution function and $R$ is the
gas constant. The chemical potential $\mu(T)$ is calculated
self-consistently such that the carrier concentration $n$ is
$T$-independent. $n$ is given by:
\begin{equation}
n=(2/V_A)\int{f(E)N(E)dE}
\label{nEQ}
\end{equation}
Where $V_A$ is the atomic volume per formula unit.

\begin{figure}
\centering
\includegraphics[width=\linewidth,clip=true,trim=0 0 0 0]{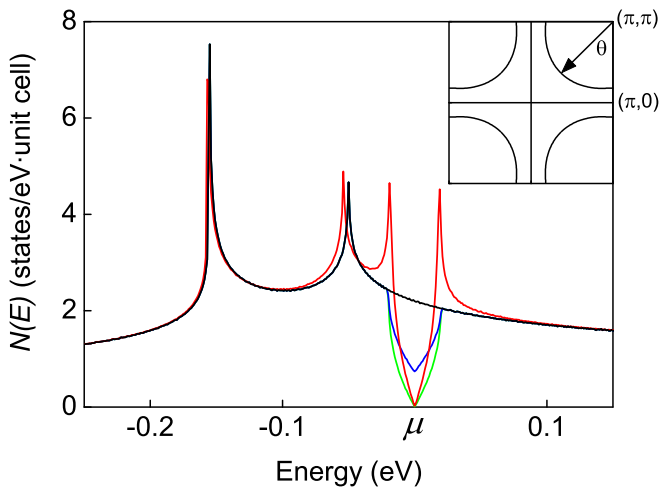}
\caption{(Color) The DOS calculated from the Bi-2212 bilayer
dispersion determined by ARPES measurements (black). Also shown is a
20meV pseudogap at 0K (green) and 100K (blue), and a 20meV SC gap at
0K (red). Inset: the Fermi surface in the $(k_{x},k_{y})$ plane
showing the angle $\theta$.} \label{DOSFIGURE}
\end{figure}

The Fermi surface (FS) in the 1st Brillouin Zone is shown in the
inset to Fig.~\ref{DOSFIGURE}. The pseudogap first forms on the FS
near $(\pi,0)$ leaving ungapped Fermi arcs\cite{FERMIARCS} between.
With decreasing temperature the Fermi arcs narrow such that the gap
seems to become nodal at $T=0$. We therefore adopt a pseudogap of
the form
\begin{equation}
E_{g}=\left\{
\begin{array}{ll}
E_{g,max}\cos{\left(\frac{2\pi\theta}{4\theta_{0}}\right)} & (\theta<\theta_{0})\\
\\
E_{g,max}\cos{\left(\frac{2\pi(\theta-\pi/2)}{4\theta_{0}}\right)} & (\theta>\frac{\pi}{2}-\theta_{0})\\
\\
0 & \mbox{otherwise}
\end{array}\right.
\label{PGEQ}
\end{equation}
where
\begin{equation}
\theta_{0}=\frac{\pi}{4}\left(1-\tanh{\left(\frac{T}{T^{*}}\right)}\right)
\label{THETA0EQ}
\end{equation}
and $T^{*}=E_{g,max}/k_{B}$.
$\theta$ is the angle shown in Fig.~\ref{DOSFIGURE}.

Eqn.~\ref{THETA0EQ} models the observed temperature dependence of
the Fermi arc length\cite{FERMIARCS2}. At $T=0$, $\theta_{0}=\pi/4$
and the pseudogap is fully nodal. As $T$ rises, $\theta_{0}$
decreases resulting in a `filling-in' of the pseudogap and the
growth of the Fermi-arcs. This model is based on results by Kanigel
\textit{et al}.\cite{FERMIARCS2} that show the Fermi-arcs collapsing
linearly as a function of $T/T^{*}$, extrapolating to zero as
$T\rightarrow0$. However we note an important feature of our model.
The Kanigel data shows the pseudogap opening abruptly at $T=T^{*}$.
A pseudogap which fills completely at $T^{*}$ would result in a jump
in the specific heat coefficient $\gamma$ at $T^{*}$, which is not
observed. The smooth evolution of the $\tanh{}$ function in Eqn.
\ref{THETA0EQ} overcomes this problem. The pseudogap is
states-non-conserving i.e. unlike the SC gap there is no pile up of
states outside the gap (see Fig. ~\ref{DOSFIGURE}). This is
implemented by eliminating states with energies $E<E_{g}$ from the
summations.

Fig.~\ref{DOSFIGURE} shows the DOS calculated from the bilayer
dispersion. The bonding and antibonding band vHs's are clearly
visible with the former 105meV below the latter. Also shown is a
20meV pseudogap at $T=0$K and 100K illustrating the gap filling with
temperature. The gap node is pinned to the chemical potential at all
$T$.

The entropy in the SC state has been modelled using a
\textit{d}-wave gap of the form
$\Delta_{\textbf{k}}=\frac{1}{2}\Delta_{0}g_{\textbf{k}}$ where
$g_{\textbf{k}}=\cos{k_{x}}-\cos{k_{y}}$. The dispersion in the
presence of the SC gap is given by
$E_{\textbf{k}}=\sqrt{\epsilon^{2}_{\textbf{k}}+\Delta^{2}_{\textbf{k}}}$
and $\Delta_{0}(T)$ is determined from the self-consistent
weak-coupling BCS gap equation\cite{PARKS}
\begin{equation}
1=\frac{V}{2}\sum_{\textbf{k}}\frac{|g_{\textbf{k}}|^{2}}{E_{\textbf{k}}}\tanh\left(\frac{E_{\textbf{k}}}{2k_{B}T}\right)
\label{BCSGAPEQ}
\end{equation}
We adopt a pairing potential of the form
$V_{\textbf{kk}^{\prime}}=Vg_{\textbf{k}}g_{\textbf{k}^{\prime}}$.
The amplitude, $V$, is assumed to be constant (=125meV) up to an
energy cut-off, $\omega_{c}$, chosen such that $T_{c}$ matches the
experimentally observed value. The pseudogap is not included in the
process of calculating $\Delta_{0}(T)$.

The superfluid density, $\rho_{s}$, is proportional to the inverse
square of the penetration depth given by\cite{RHOS}
\begin{eqnarray}
\frac{1}{\lambda_{ab}^{2}} & = & \frac{\mu_{0}e^{2}n}{4\pi\hbar^{2}}\sum_{\textbf{k}}\left[\left(\frac{\partial\epsilon_\textbf{k}}{\partial k_{x}}\right)^{2}\frac{\Delta_{\textbf{k}}^{2}}{E_{\textbf{k}}^{2}}-\frac{\partial\epsilon_{\textbf{k}}}{\partial k_{x}}\frac{\partial\Delta_{\textbf{k}}}{\partial k_{x}}\frac{\Delta_{\textbf{k}}\epsilon_{\textbf{k}}}{E_{\textbf{k}}^{2}}\right]
\nonumber\\
& \times & \left[\frac{1}{E_{\textbf{k}}}-\frac{\partial}{\partial E_{\textbf{k}}}\right]\tanh{\left(\frac{E_{\textbf{k}}}{2k_{B}T}\right)}
\label{RHOSEQ}
\end{eqnarray}
%This equation has the desirable property of explicitly yielding $\rho_{s}=0$ in %the normal limit, $\Delta_{\textbf{k}}(T=T_{c})=0$, even in the presence of a %normal-state pseudogap.
The summation in Eqs.~\ref{BCSGAPEQ} and \ref{RHOSEQ} is performed
over both the bonding and anti-bonding bands and $\Delta$ is assumed
to be the same for both bands\cite{BILAYERSCGAP}.
\begin{figure}
\centering
\includegraphics[width=\linewidth,clip=true,trim=0 0 0 0]{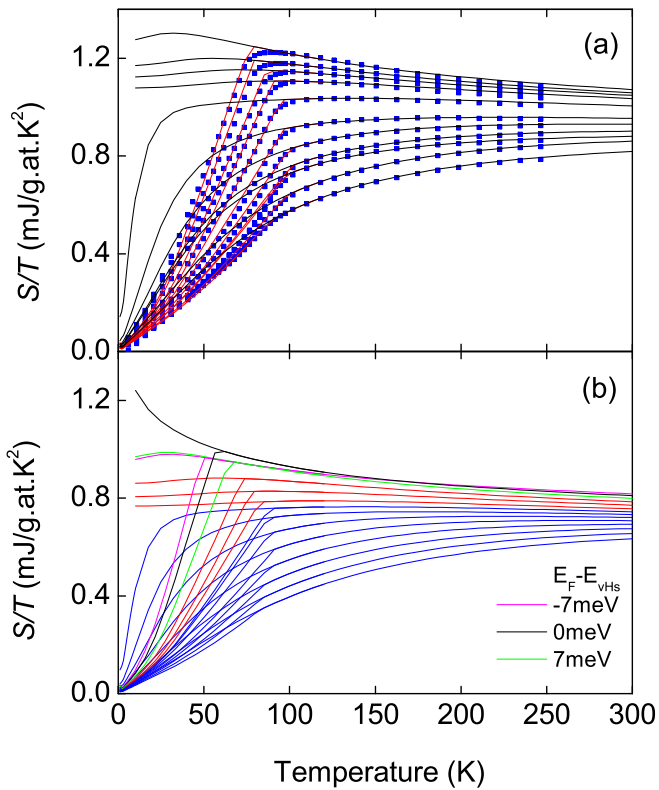}
\caption{(Color) (a) Refined normal-state (black) and SC-state (red)
fits to the electronic entropy data of Loram \textit{et
al}\cite{ENTROPYDATA2} for Bi-2212. For clarity every 20th data
point only is shown. Each curve represents a different doping level
from $p=0.129$ to 0.209. (b) Unrefined absolute entropy curves with
no fitting parameter.} \label{SOVERTFIG}
\end{figure}

The data points in Fig.~\ref{SOVERTFIG}(a) show the normal- and
SC-state entropy data of Loram \textit{et al}\cite{ENTROPYDATA2}.
Fig.~\ref{SOVERTFIG}(b) shows the absolute entropy calculated, as
described, from the dispersion with no fitting parameters. We have
merely specified the location of $E_F$ relative to $E_{vHs}$ at two
points only: in the overdoped region from Kaminski \textit{et
al}.\cite{2212VHS} and in the underdoped region from Kordyuk
\textit{et al}.\cite{UD77K} and interpolated between. The overall
$T$- and doping-dependence of the experimental data is reproduced
superbly, with absolute values just a factor of 3/4 lower. This is
our first main result. $S/T$ rises with doping and reaches its
maximum at the vHs (as observed also in
La$_{2-x}$Sr$_x$CuO$_4$\cite{ENTROPYDATA2}). A similar effect is
seen in the spin susceptibility\cite{ENTROPYDATA2} and Knight
shift\cite{SCFLUC}. In Fig.~\ref{SOVERTFIG}(a) we have rescaled the
computed entropy by the constant factor of 4/3 and refined the fit
by using $E_F$ and $E_{g,max}$ as fitting parameters. These
refinements do not alter the overall behaviour and are tightly
constrained. For example, the normal-state fits to the four most
overdoped data sets have been obtained by adjusting a single
parameter, namely $E_{F}$, as are the high-$T$ asymptotes for all
data sets. The Fermi level in the most overdoped fit is only 8meV
above the antibonding band vHs.

\begin{figure}
\centering
\includegraphics[width=\linewidth,clip=true,trim=0 0 0 0]{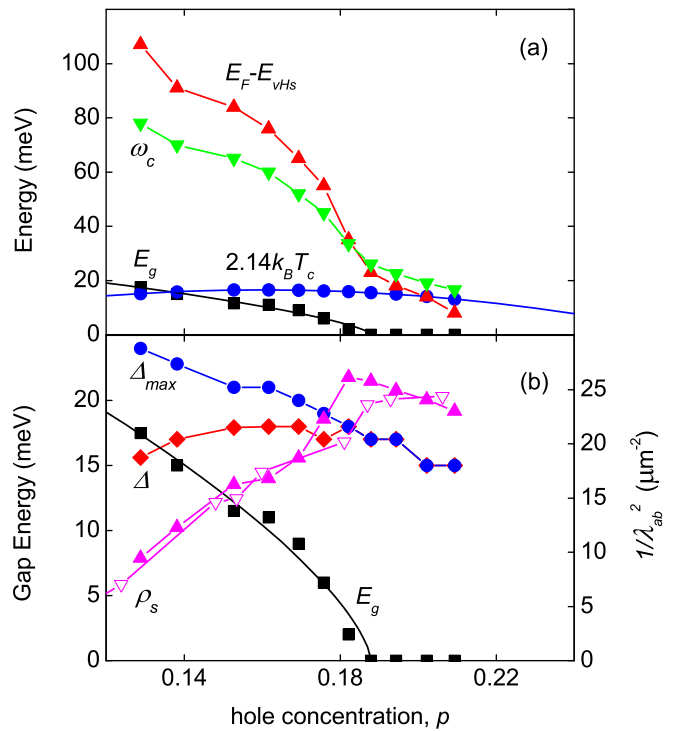}
\caption{(Color) (a) $E_{F}-E_{vHs}$ (up triangles). The vHs will
cross $E_{F}$ at $p\approx0.22$. Pairing potential energy cut-off
$\omega_{c}$ (down triangles). Pseudogap magnitude $E_{g}$ (squares)
and a fit given by Eqn.~\ref{TSTAREQ}. The measured $T_{c}$
multiplied by $2.14k_{B}$(circles). (b) Maximum gap as measured from
the DOS at 10K (circles). SC gap measured from the DOS at 10K in the
absence of a pseudogap (diamonds). Calculated superfluid density at
10K (up triangles). Measured low-$T$ Bi-2212 superfluid
density\cite{RHOS0K} (down triangles).} \label{PARAMFIG}
\end{figure}

As the doping decreases the vHs recedes from $E_{F}$ resulting in a
decrease in the number of states within $k_{B}T$ of $\mu$ and a
corresponding reduction in entropy. However as the doping is further
reduced the recession of the vHs from $E_{F}$ is no longer able to
account for the observed decrease in entropy alone and the second
adjustable parameter, the pseudogap magnitude $E_{g}$, is
introduced. This results in the progressive downturn in the normal
state $S/T$ as temperature decreases. The deduced values of
$E_{F}-E_{vHs}$ and $E_{g}$ are plotted versus doping in Fig.~
\ref{PARAMFIG}(a) along with the measured $T_c$. The doping level
has been determined from the empirical relation\cite{OCT1}
$p=0.16\pm0.11\sqrt{1-T_{c}/T_{c,max}}$. The fits suggest that the
antibonding vHs will cross $E_{F}$ near $p=0.22$ in full agreement
with recent ARPES studies\cite{2212VHS} on Bi-2212 where the
crossing occurs at $p=0.225$. This is our next key result.

The pseudogap is observed to open at critical doping
$p_{crit}=0.188$ in agreement with previous
analyses\cite{ENTROPYDATA2,OURWORK3}. $E_{g}$ has been fitted with
the following equation
\begin{equation}
T^{*}(p)=E_{g}/k_{B}=T^{*}_{0}\left(1-p/p_{crit}\right)^{1-\alpha}
\label{TSTAREQ}
\end{equation}
with $T^{*}_{0}=443.7$K and $\alpha=0.317$. These values agree with
the results of Naqib \textit{et al}.\cite{TSTAR} who determined
$T^{*}(p)$ of YBCO from transport studies. A fit to their data gives
$T^{*}_{0}=510$K and $\alpha=0.2$. The sublinear behavior of
$T^{*}(p)$ is expected if $p_{crit}$ is a quantum critical
point\cite{QCP}.

From the SC state fits the energy cutoff, $\omega_{c}$, is,
surprisingly, found to be linearly related to $E_{F}-E_{vHs}$. In particular
$\omega_{c}(meV)=10.96+0.637(E_{F}-E_{vHs})$ with correlation
coefficient $R=0.99945$. The rapid fall in $\omega_{c}$ is suggestive of magnetic or
magnetically enhanced pairing, rather than phononic.

Fig.~\ref{PARAMFIG}(b) shows the magnitude of the combined SC gap
and pseudogap, $\Delta_{max}$, measured from the calculated DOS at
10K. $\Delta_{max}$ increases with decreasing doping just as
observed from ARPES\cite{DELTAARPES},
tunnelling\cite{DELTATUNNELING} and Raman
scattering\cite{DELTARAMAN}. Also plotted is the SC gap magnitude
$\Delta$, determined by setting $E_{g}=0$ and measuring the gap in
the calculated DOS at 10K. The magnitude is smaller than typically
observed because of the weak coupling assumption for which
$2\Delta/k_{B}T_{c}=4.28$. The gap, $\Delta$, rises and falls in
conjunction with the observed $T_{c}$. Note that the experimentally
observed monotonic increase in the gap magnitude with decreasing
doping is here seen to be associated with the pseudogap, and not the
SC gap as generally believed. The behaviour here is consistent with
the two-gap picture presented by Deutscher\cite{DEUTSCHER} and more
recently by Le Tacon \textit{et al}.\cite{LETACON} but has been a
feature of our work for a long
time\cite{OURWORK1,OURWORK2,OURWORK3}.

\begin{figure}
\centering
\includegraphics[width=\linewidth,clip=true,trim=0 0 0 0]{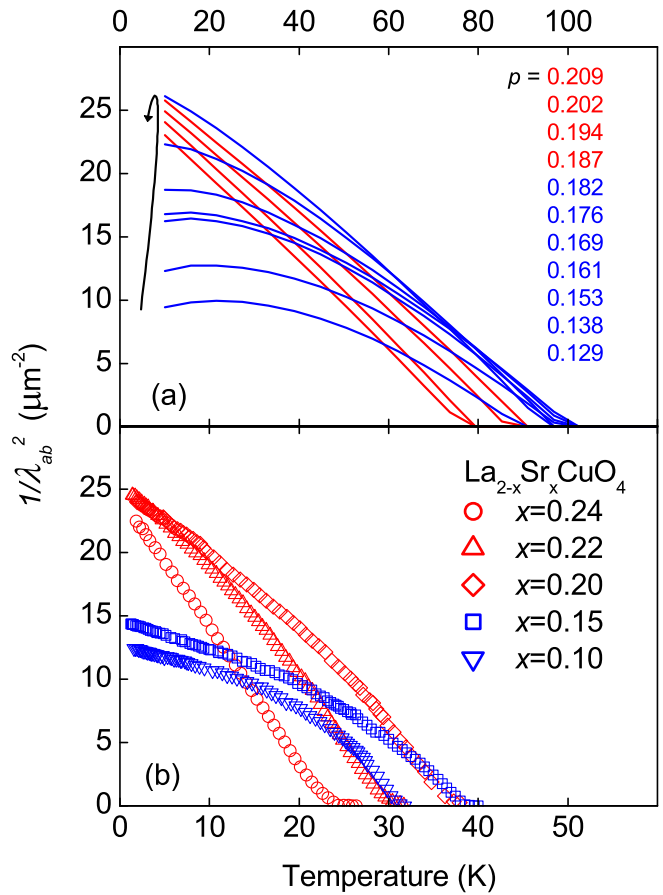}
\caption{(Color) (a) Superfluid density computed using paramaters
from the entropy fits in Fig.~ \ref{SOVERTFIG}. The arrow points in
the direction of increasing doping. Curves/data in which the
pseudogap is present are shown in blue. (b) Superfluid density of
La-214 obtained from ac-susceptibility measurements \cite{RHOSDATA}}
\label{SFD}
\end{figure}

Using the parameters obtained from the entropy fits the superfluid
density has been calculated using Eqn. \ref{RHOSEQ} with no further
adjustable parameters, and is shown in Fig.~\ref{SFD}(a). For
comparison (and in the absence of data for Bi-2212)
Fig.~\ref{SFD}(b) shows the $ab$-plane superfluid density of
La$_{2-x}$Sr$_{x}$CuO$_{4}$ (La-214) determined by ac-susceptibility
measurements\cite{RHOSDATA} on grain-aligned samples. There is
excellent agreement. The increasing linearity of $\rho_{s}(T)$ with
overdoping can now be understood in terms of the approach to the vHs
where full linearity occurs. (The crossing of the vHs in La-214 can
also be inferred from the maximum in the entropy at
$p=0.24$\cite{ENTROPYDATA2}.) The opening of the pseudogap leads to
the strong reduction in $\rho_{s}$ observed below $p=0.19$. This is
clearly illustrated by the plot of $\rho_{s}$(10K) vs $p$ in
Fig.~\ref{PARAMFIG}(b). The overall doping dependence and absolute
magnitude of $\rho_{s}$(10K) concurs almost exactly with
experimental data for Bi-2212\cite{RHOS0K} also shown in
Fig.~\ref{PARAMFIG}(b). This is our third key result. We recall that
no fitting parameters are used in Eqn.~\ref{RHOSEQ}. It is
remarkable that $S/T$ and $\rho_{s}(T)$ are so similar in La-214,
Bi-2212 and indeed Y$_{1-x}$Ca$_{x}$Ba$_{2}$Cu$_{3}$O$_{7-\delta}$,
despite the significant differences in bare band structure. The
renormalised dispersion near $E_{F}$ seems to lead to a universal
phenomenology which calls for theoretical explanation.

In the underdoped data the downturn seen at low $T$ and $p$ in the
calculated $\rho_{s}(T)$ curves arises from the closing of the Fermi
arcs and is not observed in the experimental data which show an
upturn at low $T$ and $p$. A similar downturn would occur in the
condensation energy, which again is not observed. To us this
indicates that the Fermi arc picture is, at best, incomplete. We
will discuss this elsewhere.

In summary, we have calculated the entropy and superfluid density of
Bi-2212 directly from an ARPES-derived energy-momentum dispersion.
The temperature and doping dependence of both the entropy and
superfluid density can be fully explained by the combined effects of
a proximate vHs and the opening of a normal-state pseudogap. These
results provide indirect confirmation of both the thermodynamic and
low-energy ARPES data. Fits to Bi-2212 entropy data indicate that
the antibonding vHs crosses the Fermi level near $p=0.22$ in
agreement with recent ARPES results. The superfluid density
calculated using no adjustable parameters shows excellent agreement
with experimental data and exhibits a distinctive overall
linear-in-$T$ behaviour at the vHs. The universal renormalised
phenomenology in the various cuprates, despite their differences in
bare band structure, is a key conclusion that demands theoretical
explanation. It also remains a theoretical challenge to understand
why the Fermi-liquid approach is so successful in a
strongly-correlated system.

Thanks are due to A. Kaminski for the dispersion parameters and also
Prof N.W. Ashcroft for much discussion and many helpful comments on
this work.

%%\end{multicols}
\end{document}